# Strain localization in a nanocrystalline metal: Atomic mechanisms and the effect of testing conditions

Timothy J. Rupert[1,a]

[1] Department of Mechanical and Aerospace Engineering, University of California, Irvine, California 92697, USA

[a] Author to whom correspondence should be addressed. Electronic mail: trupert@uci.edu

**Abstract:**

Molecular dynamics simulations are used to investigate strain localization in a model nanocrystalline metal. The atomic mechanisms of such catastrophic failure are first studied for two grain sizes of interest. Detailed analysis shows that the formation of a strain path across the sample width is crucial, and can be achieved entirely through grain boundary deformation or through a combination of grain boundary sliding and grain boundary dislocation emission. Pronounced mechanically-induced grain growth is also found within the strain localization region. The effects of testing conditions on strain localization are also highlighted, to understand the conditions that promote shear banding and compare these observations to metallic glass behavior. We observed that, while strain localization occurs at low temperatures and slow strain rates, a shift to more uniform plastic flow is observed when either strain rate or temperature is increased. We also explore how external sample dimensions influence strain localization, but find no size effect for the grain sizes and samples sizes studied here.

## I. INTRODUCTION

Nanocrystalline metals and alloys exhibit mechanical properties which are much improved over those of traditional microcrystalline metals, with the most notable example being higher strength.[1-4] Grain size reduction makes intragranular dislocation sources more difficult to operate, and such deformation mechanisms are even completely shut-off at the very finest grain sizes.[4] For metals with mean grain sizes ($d$) of less than ~10-20 nm, both experiments and simulations have shown that grain boundaries increasingly act as facilitators of plastic deformation through mechanisms such as dislocation emission and absorption from interfaces,[5,6] grain boundary sliding and rotation,[7,8] and grain boundary migration.[9,10] Some common features of these new mechanisms are the increased importance of the grain boundaries themselves and the highly localized nature of strain that results.

Since the disordered intercrystalline material in a grain boundary lacks the long-range order of the crystalline phase, it is possible to consider the limit of grain size refinement to be an amorphous structure. Consequently, a number of studies have uncovered interesting parallels between the mechanical response of the finest-grained nanocrystalline metals and metallic glasses. One notable feature of amorphous metals is their strength has been found to be pressure sensitive, with their properties better represented by the normal-stress dependent Mohr-Coulomb yield criterion than the von Mises criterion which describes yield in traditional metals.[11,12] Trelewicz and Schuh[13] used nanoindentation to investigate the pressure sensitivity of strength in nanocrystalline Ni-W with grain sizes from 3-100 nm and found that pressure sensitivity increased as grain size was reduced, reached a peak sensitivity at a grain size near 10 nm, and then converged to a value characteristic of metallic glasses for the very finest nanocrystalline alloys. Such behavior has been supported by atomistic modeling as well, where higher strengths were measured in compression compared to those found under tensile loading.[14] Trelewicz and Schuh[13] also investigated the strain rate sensitivity of these same nanocrystalline Ni-W alloys and found a similar trend: rate sensitivity first increased with grain refinement, reached a peak value, and then converged toward the relative rate-insensitivity observed in metallic glasses.

While strength that increases under high pressures and loading rates can be seen as advantageous for certain applications, e.g., for resisting damage during shock loading, very fine nanocrystalline metals have also demonstrated a tendency for strain localization that is reminiscent of amorphous behavior and potentially problematic for their practical application. At low temperatures and slow strain rates, metallic glasses often fail catastrophically through the formation of localized shear bands shortly after plasticity is initiated.[15-17] Recently, similar shear banding in nanocrystalline systems has been reported. Wei and coworkers observed that nanocrystalline body centered cubic (BCC) metals such as Fe,[18] Ta,[19] and W[20] fail through the formation of large shear bands when loaded in compression. Trelewicz and Schuh[13,21] studied this behavior systematically as a function of grain size in the face centered cubic (FCC) Ni-W system by using nanoindentation with a sharp cube corner tip and found that only the finest grain sizes, namely those below ~6 nm, experience strain localization. Finally, Rupert et al.[22] showed that the relaxation of nonequilibrium grain boundaries with low temperature thermal treatments, a technique that can be used to increase strength, promotes unstable plastic flow and leads to more shear banding during nanoindentation of nanocrystalline Ni-W.

MD simulations have been an invaluable tool for understanding strain localization physics in metallic glasses. For example, Cao et al.[23] found that the operation of shear transformation zones (STZs) lead to a breakdown of local icosahedral ordering in a metallic glass, causing shear band initiation. Unfortunately, such analysis has not been adequately extended to nanocrystalline materials. Without a similar atomistic understanding of the physical mechanisms behind shear localization in nanocrystalline metals, it is not clear how the plastic instability develops and grows. Shimokawa et al.[24] studied collective plasticity in nanocrystalline Al with MD simulations, but restricted their analysis to a quasi-two dimensional geometry containing columnar grains with a common out-of-plane $\langle 1\bar{1}0\rangle$ orientation (i.e., the system only contained tilt grain boundaries) and only included 8 distinct grain orientations. Sansoz and Dupont[25] used molecular statics combined with a quasi-continuum formulation to model nanoindentation and found evidence of shear localization, but also restricted their discussion to a columnar grain structure. An investigation of a three dimensional collection of nanocrystalline grains is warranted in order to accurately describe the physics of the strain localization process in a realistic nanocrystalline system.

In this paper, we perform MD simulations of nanocrystalline Ni loaded in uniaxial tension. We first uncover the atomistic mechanisms of strain localization by studying two wire samples with mean grain sizes of 3 and 6 nm, i.e., where FCC metals have been observed to experience localization in experiments. We show that localization is a process controlled solely by collective grain rotation at the smallest grain size, but dislocation activity becomes more important as grain size is increased. We then explore the effect of testing and modeling constraints such as applied strain rate, testing temperature, and sample size on localization in these materials. Our goal is to understand the conditions that promote shear banding and compare these observations to metallic glass behavior. As a whole, this work serves to provide a detailed description of the strain localization phenomenon in a nanocrystalline metal.

**II. COMPUTATIONAL METHODS**

MD simulations were performed using nanocrystalline Ni as a model system. The simulations were run with the open-source Large-scale Atomic/Molecular Massively Parallel Simulator (LAMMPS) code[26] using an integration time step of 2 fs and the embedded atom method (EAM) potential from Mishin et al.,[27] which accurately reproduces mechanical properties as well as defect energies. While the majority of prior work on nanocrystalline mechanical properties has focused on simulation cells with periodic boundary conditions, wire geometries were used here to allow for shear offsets at the free surfaces and a three dimensional nanocrystalline grain structure was simulated so that realistic localization paths could be studied. These samples were constructed by first forming nanocrystalline grain structures within a rectangular prism with periodic boundary conditions (Fig. 1(a)) using a Voronoi tessellation construction modified to enforce a minimum separation distance between grain nucleation sites, resulting in more equiaxed grains and a tighter grain size distribution. The Voronoi cells were scaled appropriately to create specimens with mean grain sizes of 3 and 6 nm and then the cylindrical wire samples were cut from these blocks, as shown in Fig. 1(b). We initially created samples with a diameter, $D$, that was four times larger than the mean grain size (i.e., $D/d = 4$) while keeping the crystallographic grain orientations constant for the two different grain sizes, but larger $d = 3$ nm samples with $D/d = 6$ and $D/d = 8$ were also created to investigate the impact of sample size. The length, $L$, of each specimen was twice the diameter (i.e., $L/D = 2$) in all cases. The simulated nanowires contained between 225,000 and 2,000,000 Ni atoms, depending on the mean grain size and the ratio of wire diameter to grain size that was chosen. Periodic boundary conditions were enforced along the wire axis, while the wire surfaces were kept free. Each wire sample was then equilibrated at 300 K and zero pressure for 100 ps using a Nose-Hoover thermo/barostat until a steady-state system energy was reached. This equilibration step is necessary for the creation of realistic nanocrystalline structures, and has been shown to reproduce experimentally measured values of sample density, grain boundary density, and excess grain boundary enthalpy.[28,29] Some specimens were deformed at 300 K, while others were cooled to 30 K over a 100 ps period for mechanical testing at low temperature.

Deformation was simulated by applying uniaxial tensile strain ($\varepsilon$) along the nanowire axis at a constant engineering strain rate while keeping zero stress on the other axes. Strain rates ($\dot{\varepsilon}$) between $5 \times 10^7$ s$^{-1}$ and $5 \times 10^9$ s$^{-1}$ were used and the temperature was held constant during testing with a Nose-Hoover thermostat. Yield strength was measured by taking the 1% offset yield stress following prior work from Brandstetter et al.[30] and Vo et al.[31] Crystal defects were visualized with the common neighbor analysis (CNA) technique, which measures the local crystal structure around an individual atom by quantifying the topology of bonds between its neighboring atoms. Figs. 1(c) and (d) show slices through the center of nanowires with $d = 6$ nm and $d = 3$ nm, respectively, where atoms are colored according to CNA. In these figures, atoms in an FCC environment are green, BCC atoms are blue, hexagonal closed packed (HCP) atoms are red, and atoms with unknown local structure appear white. Since Ni is an FCC metal in its crystalline state, grain boundaries, dislocations, and other defects appear as colors other than green. For example, a single plane of red HCP atoms represents a twin boundary while two adjacent HCP planes denote an intrinsic stacking fault. Comparison of Fig. 1(c) and (d) shows that a larger volume fraction of material is located in the grain boundaries of the $d = 3$ nm sample. Atomic-level strain tensors were computed following the work of Shimizu et al.[32] and then the local von Mises shear strain, $\eta^{Mises}$, was calculated. This quantity provided a measurement of the local inelastic deformation experienced by an individual atom. All atomistic visualization in this manuscript was performed with the open-source visualization tool OVITO.[33]

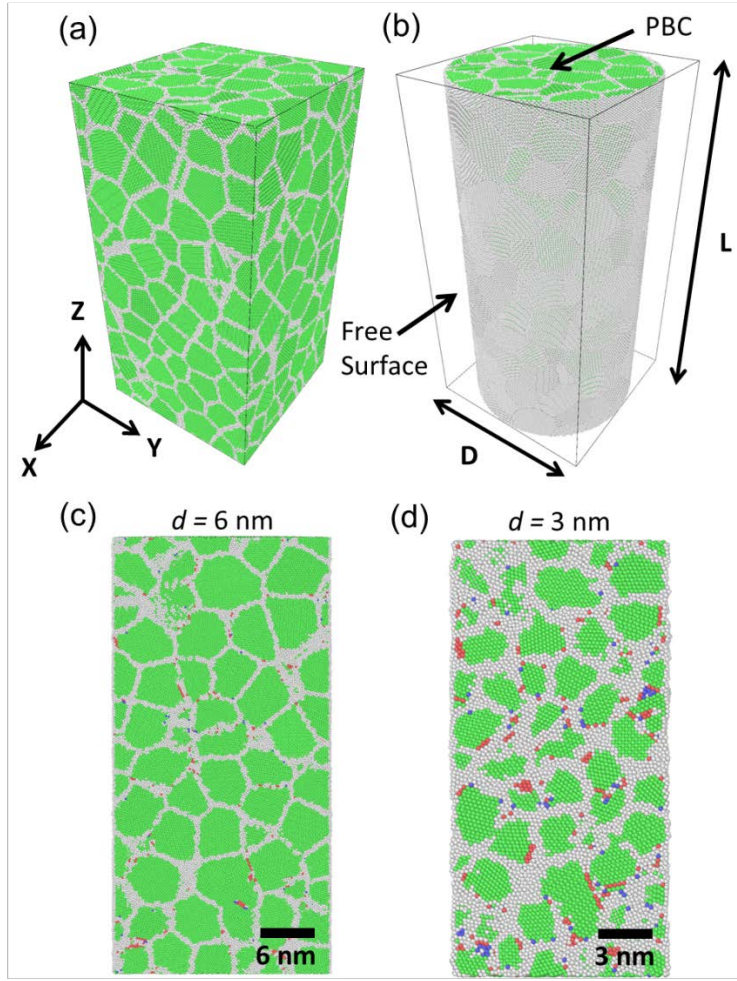

**FIG. 1.** (a) Starting grain structure for a nanocrystalline sample with $d = 6$ nm. (b) Cylindrical wire sample with $D = 24$ nm and $L = 48$ nm that was cut from the structure shown in Part (a). Slices through the center of two wire samples with $D/d = 4$ are shown in (c) $d = 6$ nm and (d) $d = 3$ nm.

## III. RESULTS AND DISCUSSION

### A. Atomic-level observations of localization processes

We first examine uniaxial tension of nanocrystalline Ni with two different grain sizes, to identify the atomic-level mechanisms that lead to strain localization. Since the metallic glass literature suggests that shear banding occurs at low temperatures and strain rates,[11,34] we began by simulating tensile deformation of our nanocrystalline wires at 30 K and with an applied strain rate of $5 \times 10^7$ s$^{-1}$ to mimic the conditions most likely to lead to highly localized plastic strain. Fig. 2(a) shows the true stress-strain curves for $d = 6$ nm and $d = 3$ nm samples tested under such conditions. The $d = 6$ nm specimen yields at a higher stress (3.9 GPa) than the $d = 3$ nm specimen (3.0 GPa), demonstrating the inverse Hall-Petch behavior that is often observed in nanocrystalline metals with grain size below ~10 nm.[13,35] However, the 6 nm grain size sample shows a strong strain softening behavior and the flow stress for this sample falls below that of the 3 nm grain size specimen for large plastic strains. The $d = 3$ nm specimen also strain softens, but in a much less pronounced manner. Both nanocrystalline wires demonstrate serrated flow, with stress drops observed since the simulated deformation is strain controlled. Similar load serrations have been observed in metallic glasses tested under constrained compression[36,37] and instrumented indentation[11,38], and these relaxation events were correlated with the formation of shear bands. A zoomed in view of the curves is presented in Fig. 2(b) to show this serrated flow more clearly and important strains are labeled for comparison with subsequent figures.

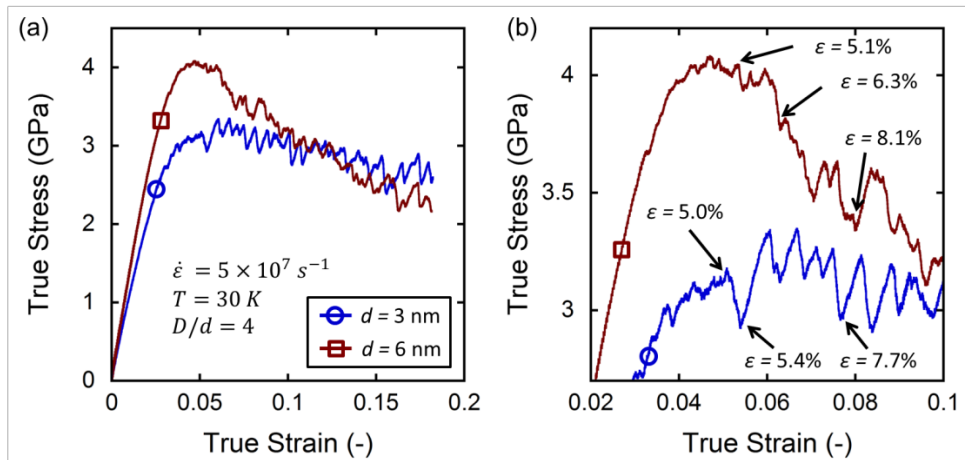

**FIG. 2.** (a) Tensile stress-strain curves for samples with mean grain sizes of 3 and 6 nm, tested at low temperature and under a slow applied strain rate. (b) Zoomed in view of the stress-strain curve showing serrated flow, with important strains marked by black arrows.

Strain localization is first explored in detail in the 6 nm grain size sample. Fig. 3 shows perspective views of the $d = 6$ nm wire at progressively larger applied strains, with atoms colored according to their local von Mises shear strain. The first frame in Fig. 3 shows the specimen during the early stages of plastic deformation, at a strain of ~4.6%. Shear strain is found predominantly at the grain boundaries, but the macroscopic deformation is still largely homogeneous. At larger applied strains, the formation of a plane with highly localized shear strain is apparent. Obvious shear offsets at the sample surface are marked with solid black arrows in the last frame of Fig. 3. By comparing the last two frames, a thickening of the strain localization region can be observed. Considering the appearance of many relaxation events in the stress-strain curve, this thickening is likely the result of multiple events occurring along the localization path. Metallic glasses demonstrate a similar behavior where the surface offsets associated with an active shear band grow with progressive straining.[23] The arrows denote the plane of most pronounced shear localization, but significant plasticity also occurs away from this region; sections near the top and in the bottom right corner also experience high shear strains. While certain metallic glasses demonstrate fully discrete plastic flow (i.e., shear banding is the only source of plastic flow), others demonstrate a combination of shear banding and homogeneous flow.[39] Schuh and Nieh[38] showed that the transition between shear banding and continuous flow during nanoindentation is dependent on loading rate, with slow loading promoting strain localization. The importance of applied strain rate on nanocrystalline strain localization will be explored in the Section B.

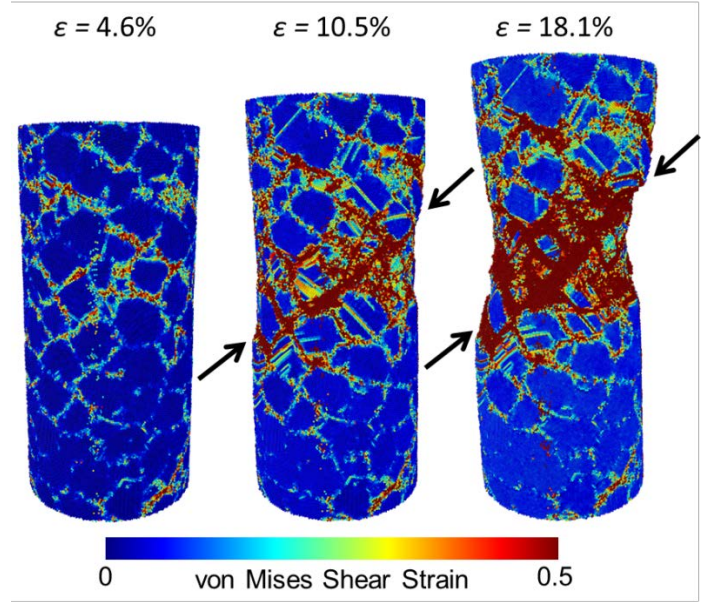

**FIG. 3. Perspective views of a *d* = 6 nm sample at progressively larger applied strains. Atoms are colored according to the local von Mises shear strain, and an obvious shear offset is denoted by solid black arrows.**

To understand the underlying mechanisms that result in strain localization, we turn our attention to the interior of the nanocrystalline wire. Fig. 4 shows a sequence of images at increasing applied axial strain, with the specimen sectioned parallel to the loading axis. Atoms are colored according to local von Mises shear strain and CNA in Fig. 4(a) and (b), respectively. During the early stages of plastic deformation shown at $\varepsilon$ = 4.6%, high shear strains occur at grain boundaries that are evenly dispersed throughout the sample. At $\varepsilon$ = 5.1%, immediately before the first major stress drop shown in Fig. 2(b), a path of high strain is percolating across the sample width. However, a grain in the center of the sample (labeled G1) is oriented such that there is no easy path for strain accommodation along its grain boundaries. A stacking fault can be seen in this grain, meaning that partial dislocation slip has occurred, but the resultant strain is not along the eventual localization path. Immediately after the first major stress relaxation, at $\varepsilon$ = 5.2%, partial dislocation slip that connects the two previously separated grain boundary paths. This intragranular slip in G1 combines with the prior grain boundary sliding to form a continuous localization path across the sample. Hasnaoui et al.[40] and Sansoz and Dupont[25] observed a similar formation of shear planes through combined grain boundary sliding and intragranular slip during when simulating tension at elevated temperatures and nanoindentation, respectively. However, these authors did not explore how the formation of such paths influenced subsequent plastic deformation.

As applied strain increases beyond 5.2%, the strain localization intensifies along the shear plane. This appears as a shift to dark red coloring ($\eta^{Mises} > 0.5$) along the grain boundaries and a thickening of the grain boundary strain path. The stacking fault becomes a twin boundary and the twin in G1 thickens as a result of multiple partial dislocations moving through the grain. Fig. 4(b) shows that stored stacking faults and deformation twins are primarily observed within the strain localization region. While partial dislocation motion provides the intragranular contribution of the localization path here, other nanocrystalline metals could mimic this mechanism with successive full dislocation slip or combinations of leading and trailing partial dislocations. Secondary shear bands form parallel and perpendicular to the original conduit in the middle of the sample, but the highest strains occur along the initial path. To highlight the fact that strain localizes once the path crosses the sample, attention is drawn to the bottom right of the sample in Fig. 4(a). A small region of high strain develops during the early stages of plastic deformation ($\varepsilon = 4.6\%$), but this region does not grow and the strain even relaxes slightly during the remainder of the experiment.

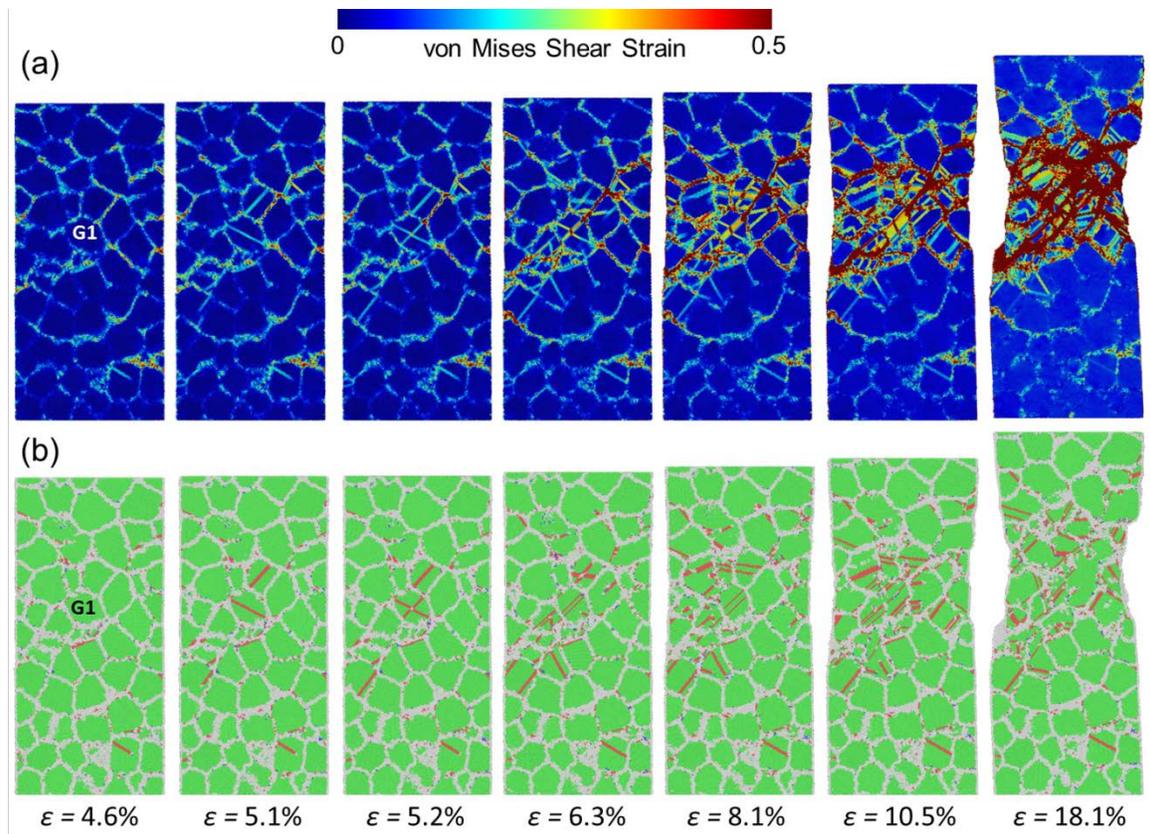

**FIG. 4. Sequence of images taken from a slice of the *d* = 6 nm wire. Atoms are colored according to local von Mises shear strain in (a) and according to CNA in (b). G1 denotes a grain that undergoes deformation twinning during the simulation.**

Fig. 5 presents zoomed images of the twinned grain to show the deformation twinning process in further detail. Figs. 5(a) and (b) show the grain before and after the first partial dislocation slip that completes the localization path. The twin forms and grows through a progressive migration process highlighted in Figs. 5(c)-(e). A partial dislocation travels part of the way across the grain before stopping. This incomplete migration is observed repeatedly in these simulations, with the stopping point always corresponding to atoms involved in the intersection of the two stacking faults shown in Fig. 4(b). Eventually, with additional time and strain, the partial dislocation is pushed all the way across the grain, causing the twin to grow by one lattice plane. The twin continues to grow with progressive straining on planes both above and below the original twin. By comparing Figs. 5(a) and (f), one can see that the deformation twinning mechanism produces a large shear strain in the grain of interest; the grain experiences a simple shear strain of ~30% between these two states. A fiducial {111} plane is marked with a black line to highlight the resultant deformation of the grain.

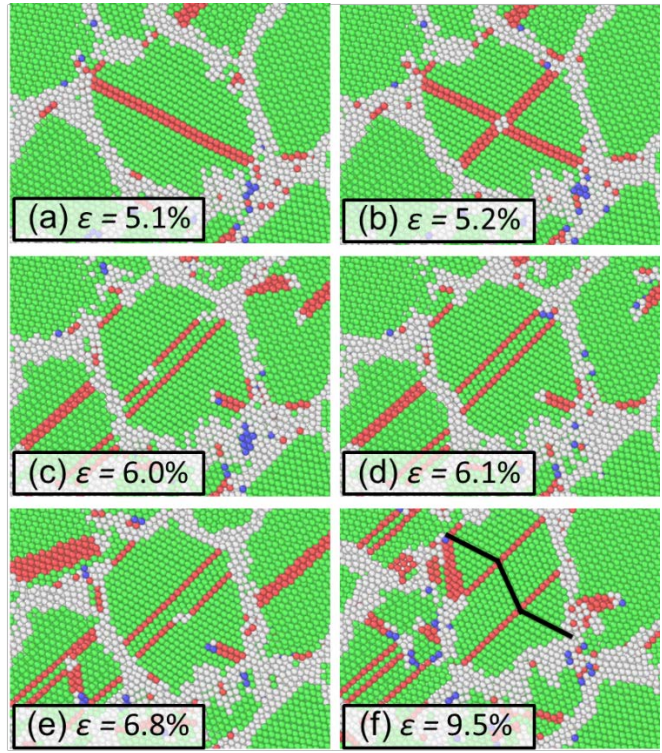

**FIG. 5. Details of the deformation twinning process. (a) Before the localization path is complete. (b) After partial dislocation slip completes the localization path by connecting two sliding grain boundaries. A twin forms and grows in (c)-(f). The black line in (f) marks a fiducial {111} plane.**

Significant grain coarsening is observed in the region of highly localized strain. Fig. 6 highlights this by showing the grain structure before testing and after an applied strain of 18.1%. Within the highly strained region, marked with dotted black lines, a number of grains larger than the as-prepared grain size are found. Two clear examples are denoted with asterisks; these grains are also elongated. Away from the strain localization, grains remain equiaxed and a similar size as the starting structure. Mechanically-induced grain growth has been observed in a number of nanocrystalline metals such as Al,[10,41] Ni,[42,43] and Cu,[9] as well as alloys such as Ni-Fe,[44,45] Ni-W,[46,47] and Co-P.[48] This grain growth can be caused by a combination of grain boundary migration and coalescence due to grain rotation, and has been shown to be driven by high shear stress.[49,50]

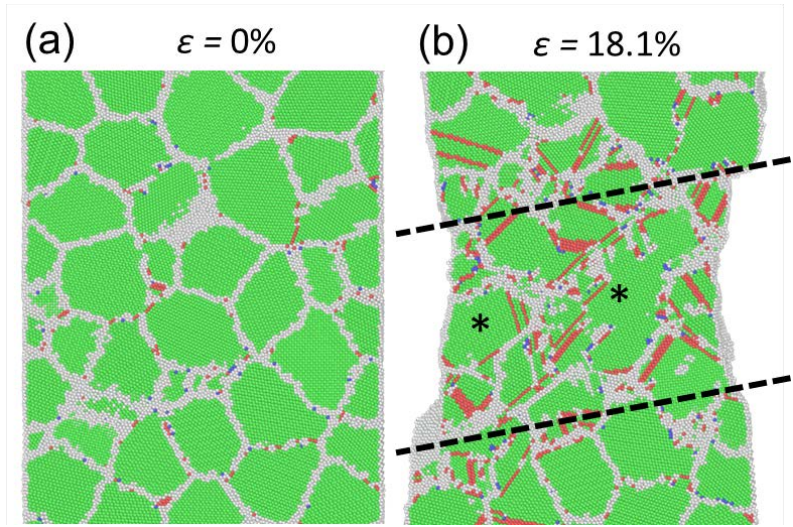

FIG. 6. The grain structure of the $d$ = 6 nm sample, colored according to CNA, is shown (a) before straining and (b) after 18.1% applied strain. Two coarsened and elongated grains are marked with asterisks, while the strain localization region is marked with dashed black lines.

We next investigate strain localization mechanisms in the $d = 3$ nm specimen. Images of this sample, sliced to view the interior, are shown in Fig. 7 for various applied strains. Atoms are colored according to local von Mises shear strain and CNA in Fig. 7(a) and (b), respectively. While small areas of high shear strain are observed at $\varepsilon = 5.0\%$, a path of high strain that spans the sample becomes clear at approximately $\varepsilon = 5.4\%$, after a major stress drop in the stress-strain curve presents in Fig. 2(b). This localization path is created through percolation of high strain in grain boundary regions only, suggesting that localization at the finest grain sizes is controlled solely by interfacial mechanisms such as grain boundary sliding and grain rotation. A smaller grain size means that a larger volume fraction of the sample is grain boundary material, making it easier to find a continuous interfacial route across the sample. Grain boundary dislocation mechanisms are also suppressed as grain size is reduced,[51,52] making it harder for a step-wise deformation twinning mechanism to operate. The strain along this path intensifies with progressive straining, with the colors in Fig. 7(a) shifting from light blue and yellow ($\eta^{Mises}$ ~ 0.2-0.3) to dark red ($\eta^{Mises} > 0.5$). Stored stacking faults are found after large plastic straining, but we do not observe the progressive deformation twinning mechanism observed in the 6nm grain size sample.

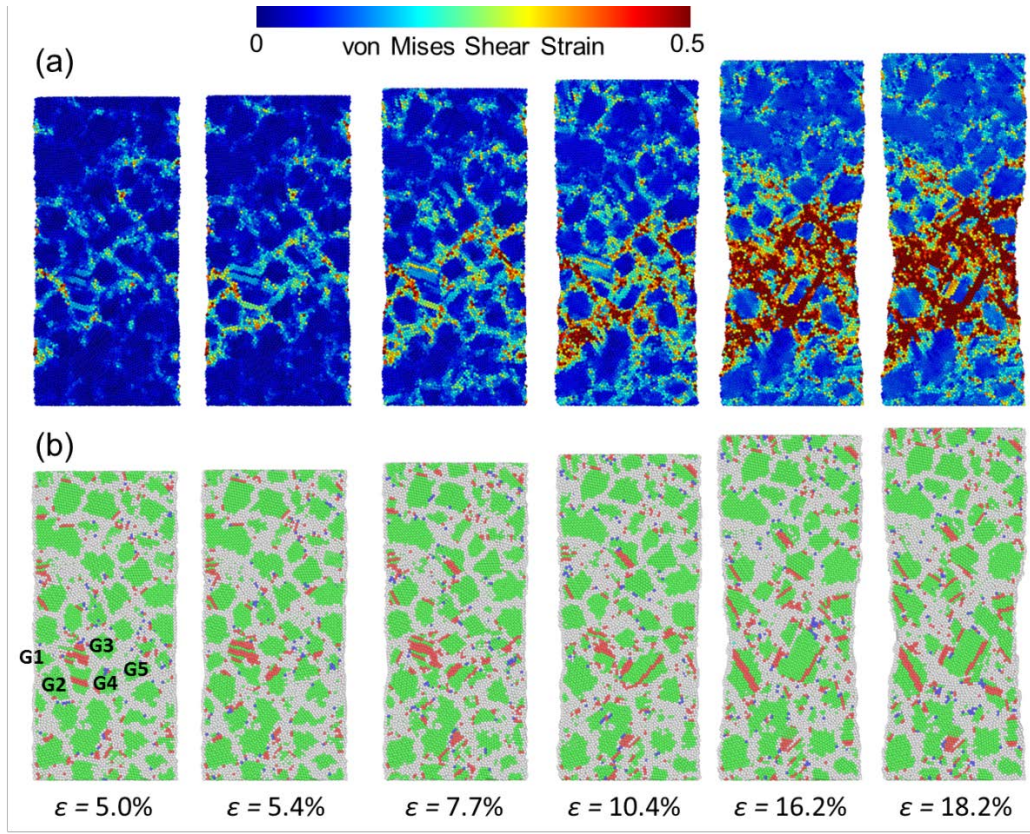

**FIG. 7. Sequence of images taken from a slice of the $d = 3$ nm wire. Atoms are colored according to local von Mises shear strain in (a) and according to CNA in (b). The grains denoted as G1-G5 in (b) coalesce through rotation and sliding to form two larger grains by the end of the test.**

Within the highly strained region, a number of grains are found to coalesce to form larger crystallites. For example, in Fig. 7, the grains G1 and G2 rotate and slide until they find a common orientation, as do grains G3, G4, and G5. The mechanically-induced grain growth found in the $d = 3$ nm sample is further highlighted in Fig. 8 for a different slice of the sample.

Fig. 8(a) and (b) show the structure colored according to CNA before testing and after 18.2% applied strain, respectively, while Fig. 8(c) shows the deformed configuration with atoms colored according to $\eta^{Mises}$ to highlight the strain localization. Within the localization region, a number of coarsened grains can be found. Three specific examples are marked with asterisks. Again, similar to the results shown in Fig. 6 for the larger grain size sample, the grain size remains stable away from this highly strained region. Near the top of the sample, where plastic strain is low, the grain size is similar to the as-prepared structure.

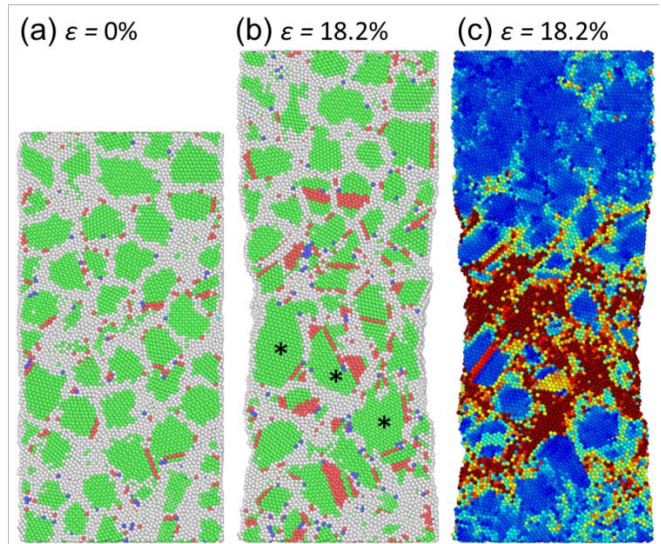

**FIG. 8.** The grain structure of the $d = 3$ nm sample, colored according to CNA, is shown (a) before straining and (b) after 18.2% applied strain. Three coarsened grains are marked with asterisks in (b). Atoms are colored according to their local von Mises shear strain in (c), with the same coloring scheme as Figs. 3, 4, and 7.

For both grain sizes, we find that the formation of an easy shear path across the sample is a prerequisite for strain localization. This path can be formed through a combination of grain boundary and dislocation mechanisms ($d = 6$ nm), or entirely through grain boundary plasticity ($d = 3$ nm). Once the strain path is formed, it is followed by rapid strain localization along that same plane as well as significant stress-driven grain growth. Without clear hardening mechanisms such as intragranular dislocation tangling and storage, there is nothing to stop runaway localization. These nanocrystalline localization mechanisms are inherently different from the collective STZ operation that leads to shear banding in metallic glasses, as grain size and structure remains an important factor limiting the localization path. While a shear transformation zone is a transient event and can occur anywhere, grain sliding and rotation paths are limited by the connectivity of the interfacial network. Based on the observation above, we suggest two potential methods for suppressing strain localization: (1) selective doping to resist grain coarsening and sliding and (2) breaking up the grain boundary percolation path. Both experimental and computational research has shown that grain boundary migration and sliding are restricted if interfaces are doped with impurities such as H and O,[53-55] or other metals such as Nb and Fe.[56-58] Careful doping of nanocrystalline metals should delay localization until higher applied stresses, but may not remove the problem altogether. Alternatively, the importance of an interfacial path for sliding suggests that grain boundary engineering, the planned alteration of the interfacial network topology and character, could be an effective way of avoid localization. Low energy boundaries (referred to as "special" in the materials science literature) would be particularly resistant to sliding and migration, and could limit the percolation path for strain localization if added judiciously.

## B. Effect of testing conditions on strain localization

Having explored the atomistic mechanisms behind nanocrystalline strain localization, we now turn our attention to understanding its phenomenology. Shear banding in metallic glasses is known to depend strongly on testing conditions such as strain rate and temperature. Spaepen[59] introduced the first deformation maps for metallic glasses, which delineated between different spatial distributions of plastic strain during deformation. Homogeneous and inhomogeneous regimes were found at high and low temperatures, respectively.  For an amorphous metal, homogeneous deformation results from the gradual emergence of viscous flow as temperature increases.  Schuh et al.[34] expanded such deformation maps to include a second inhomogeneous-to-homogeneous transition that depends on strain rate and incorporates the collective dynamics of STZs during shear band nucleation.  If the phenomenology of nanocrystalline localization is similar to that of a metallic glass, one would expect a shift from the strain localization observed above in Section A (slow strain rate of $5 \times 10^7$ s$^{-1}$ and low testing temperature of 30 K) to more homogeneous plasticity as strain rate or temperature are increased while other testing conditions remain constant.

We begin by exploring the effects of strain rate on plastic localization, while keeping testing temperature constant at 30 K.  Fig. 9 presents uniaxial stress-strain curves for different engineering strain rates, with the 6 nm grain size shown in Part (a) and the 3 nm grain size shown in Part (b).  As strain rate is increased, the flow serrations in the stress-strain curve become less pronounced.  Increasing strain rate to $5 \times 10^8$ s$^{-1}$ only causes a small increase in yield strength of ~0.1 GPa for both grain sizes.  However, further increasing the strain rate to $5 \times 10^9$ s$^{-1}$ leads to a much larger increase in yield strength and flow stress (instantaneous yield strength or the stress required to continue plastic flow) for both samples.  Brandl et al.[60] studied nanocrystalline Ni with $d$ = 11.5 nm using MD and observed a temporary overshoot in the stress-strain curve as strain rate was increased to high values.  However, the overshoot that these authors observed diminished with increasing plastic strain and was attributed to a delay in dislocation propagation at high strain rates.  On the other hand, our stress-strain curves have a consistent shape but are shift upward by a constant value for the entire plastic regime when $\dot{\varepsilon} = 5 \times 10^9$ s$^{-1}$.  The fastest $d$

= 6 nm curve is shifted upwards by ~1 GPa and the fastest $d = 3$ nm curve is ~0.5 GPa higher, when compared to the slowest applied strain rate.

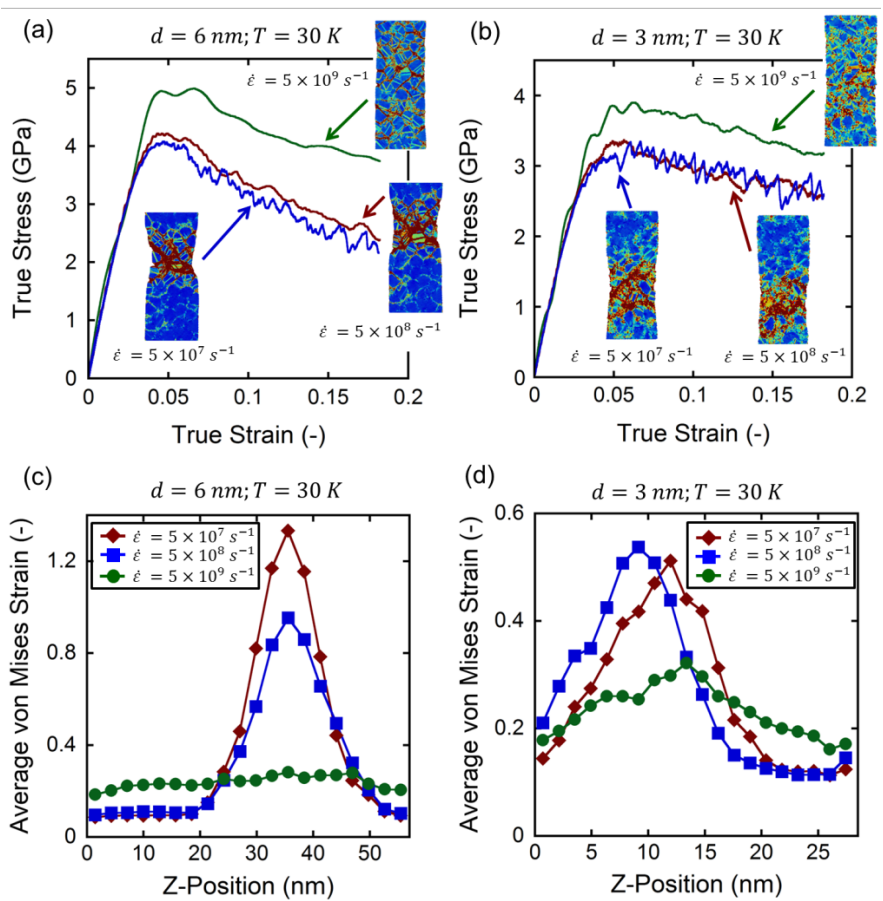

**FIG. 9. Tensile stress-strain curves for (a) 6 nm grain size and (b) 3 nm grain size samples tested at different strain rates, while temperature is kept constant at 30 K. Inset to (a) and (b) are atomic configurations taken at the end of the tension simulations with atoms colored according to local von Mises shear strain. The average von Mises shear strain is presented as a function of position along the wire length in (c) and (d).**

Images with atoms colored according to local von Mises strain are inset in Figs. 9(a) and (b) to show the spatial distribution of plastic strain in each sample; the atomic configurations are all taken from the end of the tensile experiment when $\varepsilon = 18.2\%$. For both grain sizes, the plastic strain in the two slower strain rate simulations appears to be strongly localized while it is much more homogeneously distributed throughout the length of the sample at the highest strain rate. To improve upon these visual cues, we divide each wire sample into 20 pieces along its length (i.e., the loading axis or z-axis) and plot the average von Mises shear strain in Figs. 9(c) and (d) as a function of position. Focusing first on the 6 nm grain size samples shown in Fig. 9(c), a sharp peak in the average von Mises shear strain is observed near the center of the sample for the slowest strain rate, highlighting the spatial localization of plasticity. The height of this peak decreases as strain rate increases to the intermediate value, and then the strain profile appears completely flat for the fastest strain rate. For the 3 nm grain size in Fig. 9(d), a peak of approximately the same height and width is observed for the slowest and the intermediate strain rates, suggesting that they experience a comparable level of localization. For the fastest strain rate, the strain profile again begins to flatten out, although not completely as some localization persists. Although the strain rate needed to completely suppress localization appears to depend on grain size, a consistent trend of faster strain rates causing a transition from discrete yielding to continuous flow is observed. At high strain rates, a single shear localization event cannot keep up with the applied strain and many plastic events are needed, leading to a more uniform spatial distribution of strain. While higher strain rate should lead to some subtle strengthening, the suppression of catastrophic shear banding may be responsible for the exaggerated strengthening observed at $\dot{\varepsilon} = 5 \times 10^9$ s$^{-1}$. As a whole, the strain rate dependence of strain localization in nanocrystalline Ni appears to be similar to that observed for shear banding in metallic glasses.

We next explore the effect of temperature on nanocrystalline strain localization by running additional simulations at 300 K. The simulations were run at $\dot{\varepsilon} = 5 \times 10^8$ s$^{-1}$ since this strain rate leads to localization at low temperature, but requires less simulation time. Stress-strain curves for the 6 nm and 3 nm grain sizes are presented in Figs. 10(a) and (b) respectively. Increasing testing temperature to 300 K leads to lower strengths, and there is also less strain softening for both grain sizes. The number of serrations in the stress-strain curve decreases as temperature is increased for $d = 3$ nm, but it is difficult to make any such statements about the curves for $d = 6$ nm. Images of the sample colored according to local von Mises strain are included as insets to both figures. While some degree of localization is still present during testing at 300 K, the strain away from this localization region increases and the strain distribution becomes more homogeneous. This can be seen more clearly in Fig. 10(c), where we slice the sample into 20 pieces and plot the average von Mises strain for these sections as a function of their position. To compare the two grain sizes on the same graph, we normalize the z-position by the length of the wire. For both samples, increasing temperature reduces the height and increases the width of the localization peak, leading to a flatter strain distribution. Again, the transition to more spatially homogeneous strain at elevated temperatures mimics metallic glass behavior. It is expected that even higher testing temperatures would lead to further homogenization of the plastic strain. However, higher temperatures would be above our equilibration treatment temperature and could cause thermal grain growth that would complicate a direct comparison, so we do not perform such simulations here.

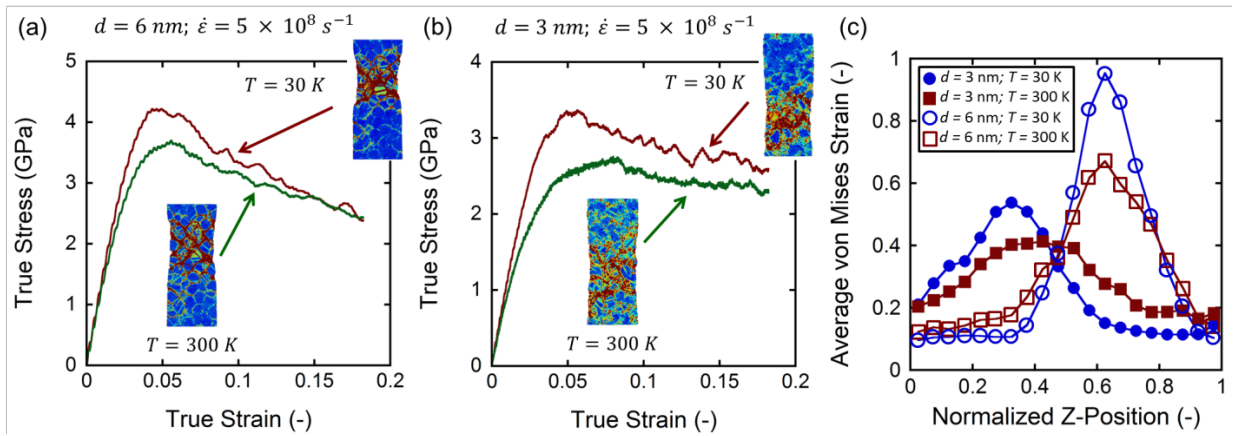

**FIG. 10.** Tensile stress-strain curves for (a) 6 nm grain size and (b) 3 nm grain size samples tested at different temperatures, while strain rate is kept constant at $5 \times 10^8$ s$^{-1}$. Inset to (a) and (b) are atomic configurations taken at the end of the tension simulations with atoms colored according to local von Mises shear strain. The average von Mises shear strain is presented as a function of normalized position along the length of the wire in (c).

Finally, we investigate the effect of sample size on strain localization by simulating 3 nm grain size samples with different wire diameters and, therefore, different numbers of grains through the sample thickness. Experimental evidence suggests that the mechanical behavior of nanocrystalline pillar/wire samples can be altered if the characteristic extrinsic length scale of the experiment (i.e., the sample dimensions) becomes comparable to the characteristic intrinsic length scale of the material (i.e., the grain size), although some reports suggest a softening effect with decreasing sample size[61,62] while others report strengthening under similar conditions.[63] Recent MD simulations from Zhu et al. have shown that these conflicting size scaling trends are both possible, with grain size determining which trend is observed.[64] These authors found that a larger grain size of 20 nm experienced softening as external dimensions were reduced while a smaller grain size of 5 nm experienced strengthening. In addition, we always observe mechanically-induced grain growth with our strain localization here, but such behavior could also be influence by sample size. Using MD simulations of thin film geometries, Gianola et al.[55] found that mechanically-induced grain coarsening is substantially enhanced near free surfaces and multiple authors showed that this surface effect occurs over a length that is roughly the order of the grain size.[55,65] Our goal here is to check that our observations are not an artifact of our relatively limited sample size.

Additional $d$ = 3 nm wires were created with diameters of 18 and 24 nm, to complement our original sample with $D$ = 12 nm. Therefore, we tested samples with $D/d$ ratios of 4, 6, and 8, and the largest sample had ~2,000,000 atoms and ~1550 grains. Tensile simulations were run at a strain rate of $5 \times 10^8$ s$^{-1}$ and a temperature of 30 K. Fig. 11(a) shows stress-strain curves from these three samples. Yield strength was unaffected by sample size here (3.0 GPa for all three samples), although the number and severity of flow serrations decreasing as sample size becomes larger. Fig. 11(b) presents images were atoms are colored according to $\eta^{Mises}$ for $\varepsilon = 18.2\%$, and spatial strain localization is observed for all three samples. This is confirmed by Fig. 11(c), where the average von Mises strain is plotted against normalized z-position and all samples show a peak in average strain. The mechanical properties and spatial distribution of strain do not appear to be affected by external sample size for these simulations.

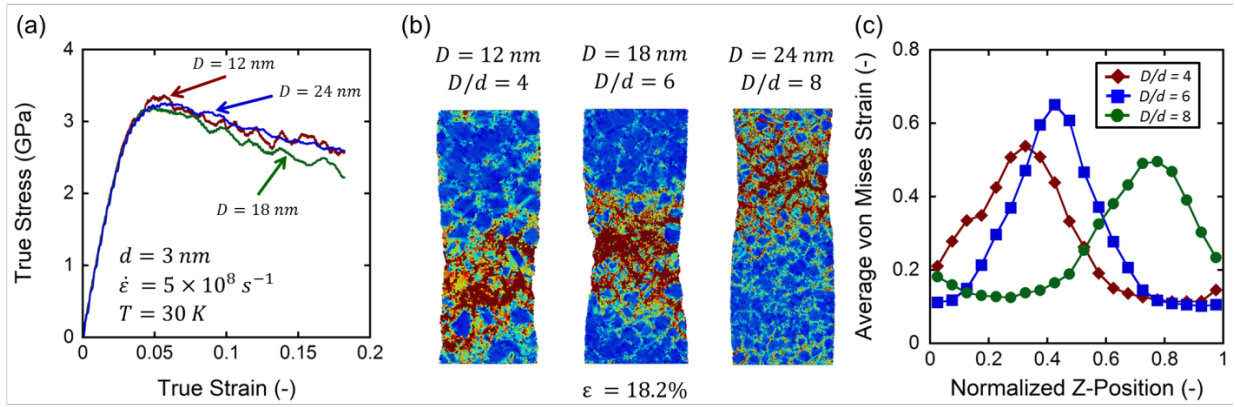

**FIG. 11.** (a) Tensile stress-strain curves for 3 nm grain size samples with different wire diameters, while strain rate and temperature are kept constant at $5 \times 10^8$ s$^{-1}$ and 30 K, respectively. (b) Atomic configurations taken at the end of the tension simulations with atoms colored according to local von Mises shear strain. The average von Mises shear strain is presented as a function of normalized position along the length of the wire in (c).

We also investigate the interior grain structure after localization to ensure that the mechanically-induced grain growth observed in Figs. 6 and 8 was not an artifact caused by small sample size. Figs. 12(a) and (b) show the largest wire ($D/d = 8$) which has been cut down the middle and colored according to CNA for 0% and 18.2% strain, respectively. In Part (b), the strain localization region is denoted by dotted black lines. As we also observed for the smaller diameter samples, rampant grain growth is found in the highly strained region while the areas away from this region at the bottom of the wire show a grain structure and size which is reminiscent of the undeformed wire. Coarsened grains can be seen at the very center of the sample, suggesting that the growth we see here cannot solely be caused by proximity to a free surface but rather by the high plastic strains in the localization zone.

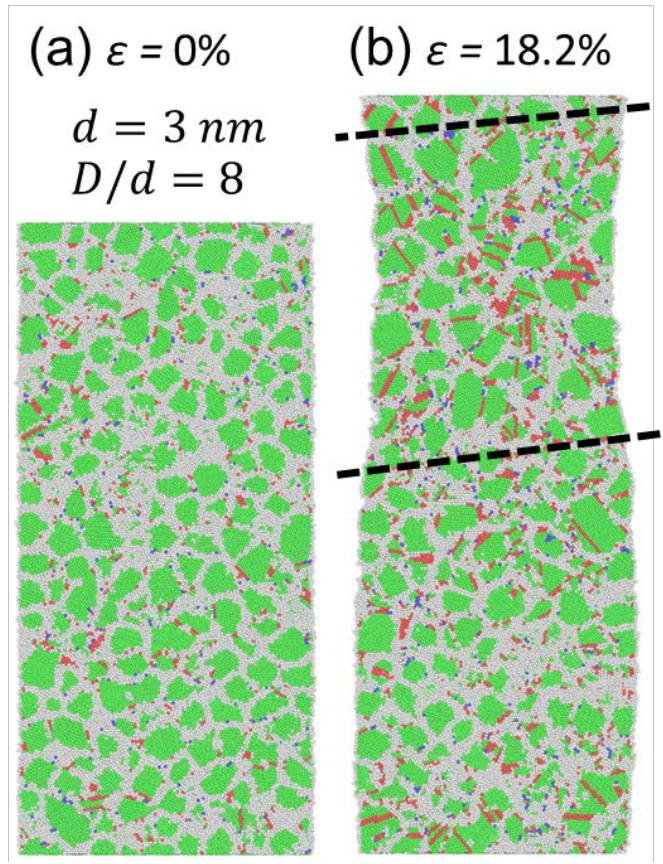

**FIG. 12. The grain structure of the largest $d = 3$ nm sample ($D/d = 8$), colored according to CNA, is shown (a) before straining and (b) after 18.2% applied strain. Coarsened grains are observed in the strain localization region marked by dashed black lines, even near the middle of the wire away from the surfaces.**

## IV. CONCLUSIONS

In this article, we have used MD simulations to study strain localization in a model nanocrystalline metal. The results presented here provide insight into the atomic mechanisms responsible for catastrophic yielding in nanocrystalline Ni, while also highlighting the importance of testing conditions on such localization. The following conclusions can be drawn:

- Strain localization occurs when a high strain path percolates across the sample width. This path can be formed entirely along grain boundaries or through a combination of grain boundaries and intragranular dislocation motion. For a 6 nm grain size, deformation twinning caused by successive partial dislocation motion can extend this localization path through a crystal interior.

- Mechanically-induced grain growth was observed in the strain localization region for both grain sizes probed here. However, away from this region of high plastic strain, the grain structure is unaffected by the applied loading.

- While strong strain localization is found when testing is carried out at slow strain rates and low temperatures, a shift to more uniform plastic flow is observed when strain rate or temperature is increased. These trends mimic the phenomenology of shear banding in metallic glasses, suggesting a similarity in deformation physics with both exhibiting collective plasticity.

- Sample size was not found to noticeably impact yield strength, degree of strain localization, or grain coarsening in our simulations, meaning the behavior we observe here should translate to larger nanocrystalline samples.

The results presented here provide a physical explanation for a catastrophic failure mode that has been observed in experimental testing of nanocrystalline metals. By showing how this plastic instability develops and grows across nanocrystalline samples, we hope to enable the development of strategies for avoiding strain localization in these materials. Specifically, our results suggest that careful doping and grain boundary network engineering may be promising approaches for the suppression of strain localization. The work presented here also provides another connection between nanocrystalline and amorphous mechanical behavior, supporting the ideas that these materials exist on a structural continuum and that their deformation physics are similar.

**ACKNOWLEDGEMENTS**

This work was supported by the Broadening Participation Research Initiation Grants in Engineering (BRIGE) program from the National Science Foundation under Grant No. (CMMI-1227759).